\documentclass[12pt]{iopart}
\pdfoutput=1
\usepackage{iopams}  
 \expandafter\let\csname equation*\endcsname\relax
  \expandafter\let\csname endequation*\endcsname\relax
\usepackage{amsmath}
\usepackage{amsthm}
\usepackage{amsfonts}
\usepackage{amssymb}
\usepackage{graphicx}
\usepackage{cmap}
\usepackage{xcolor}
\usepackage[T1]{fontenc}
\usepackage[utf8]{inputenc}
\usepackage[english]{babel}

\newtheorem{thm}{Theorem}
\newtheorem{lem}{Lemma}

\newtheorem{rem}{Remark}
\newcommand{\la}{\lambda}
\newcommand{\mla}{\langle \lambda \rangle}
\newcommand{\bb}[1]{\mathbb{ #1 }}

\newcommand{\beq}{\begin{equation}}
\newcommand{\eeq}{\end{equation}}

\def\XXint#1#2#3{{\setbox0=\hbox{$#1{#2#3}{\int}$}
     \vcenter{\hbox{$#2#3$}}\kern-.5\wd0}}

\begin{document}

\title[Asymptotic analysis of fitness maximization with noise]{Asymptotic analysis of noisy fitness maximization, applied to metabolism \& growth}

\author{Daniele De Martino$^1$ \& Davide Masoero$^2$}

\address{$^1$ Institute of Science and Technology Austria (IST Austria), Am Campus 1, Klosterneuburg A-3400, Austria \\
$^2$ Grupo de Física Matemática da Universidade de Lisboa, Alameda da Universidade, 1649-004 Lisboa, Portugal}
\vspace{10pt}


\begin{abstract}
We consider a  population dynamics model coupling cell growth  to a diffusion in the space of metabolic phenotypes as it
can be obtained from realistic  constraints-based modeling.
In the asymptotic regime of slow diffusion, that coincides with the relevant experimental range, the resulting non-linear Fokker-Planck equation
is solved for the steady state in the WKB approximation
that maps it into the ground state of a quantum particle in an Airy potential plus a centrifugal term. We retrieve scaling laws for growth rate fluctuations and time response with respect to the distance from the maximum
growth rate suggesting that suboptimal populations can {have a faster  response to perturbations}.
\end{abstract}

\section*{Introduction}
The problem of statistical mechanics of studying the emerging macroscopic behavior of a system from the microscopic dynamics of its
interacting units has been successfully solved in many condensed matter issues
but its concepts and techniques reach beyond physics. Indeed it is of the greatest interest to apply the methods of statistical physics to molecular biology\cite{bialek2012biophysics, sneppen2005physics}. The inherent complexity of many biological phenomena  requires a faithful integration of many  microscopic biochemical mechanisms and in turn an analysis of the interplay between noise  and interactions that could borrow from the tools of statistical mechanics. Examples range  from the study of structural changes of polymers in solutions\cite{de1979scaling} in terms of self-interacting random walks to modeling brain functions and neural networks with disordered spin models\cite{amit1992modeling}.      
In particular, the process of cell growth can be in principle studied by analysing the dynamics of a large network of enzymatic reactions (metabolism) where the growth rate becomes a function of the metabolic state by simple stoichiometric calculations. 
The simplest choice of a metabolic state that maximises the growth, based on the exponential taking over of the fastest phenotype, has lead to
qualitative and semi-quantitative predictions of the metabolic state and its response to genetic knock-outs for stationary
growing microbial cultures, a framework that has been called flux balance analysis (FBA)\cite{ibarra2002escherichia}.
On the other hand, a simple experimental observation is that  growth rates fluctuate from cell to cell in the same
population even for stationary cultures of identical (isogenic) cells. 
We remark that recent advancements in tracking techniques and/or keeping stationary conditions in growing cultures\cite{wang2010robust, ullman2013high} have lead to the possibility of measuring single cell growth rates and division times distributions with unprecedented
precision and stability raising a surge of interest about the laws ruling growth rate variability, that are known to be connected 
with many general biological issues like ageing\cite{wang2010robust}, size homoeostasis\cite{taheri2015cell}, the quest
for mechanisms that trigger cell division\cite{kennard2016individuality} and noise in gene expression levels\cite{shahrezaei2015connecting}, in particular of metabolic enzymes\cite{kiviet2014stochasticity}.
From the point of view of modelling, and in regards to the connections with the underlying metabolic dynamics, an understanding of such fluctuations would require to extend the FBA approach  that by definition retrieves a single value and no fluctuations.
An extension of the FBA approach in order to incorporate the fluctuations that naturally arise taking into consideration an underlying microscopic dynamics  was recently proposed in \cite{de2016growth}
where both a  max-entropy approach (i.e. based on fixing the average growth rate in the metabolic space) and a simple
diffusive model have been put forward.   
The former revealed  that fast-growing phenotypes occupy a very low-entropy region of the metabolic space and from a
biological point of view this would imply that  enzyme  expression levels shall be highly fine tuned in this region,
possibly overcoming   shot noise limits\cite{berg1977physics}. 
The latter leads to the following non-linear Fokker-Planck equation for the evolution of the
probability distribution $p(\la)$ of growth-rates or fitness $\la$
\begin{equation}
\dot{p}(\lambda) = (\la-\mla)
p(\la) +D \left[\frac{\partial^2 p}{\partial\la^2} -\frac{\partial}{\partial\la} \left[p(\la)
\frac{\partial}{\partial\lambda}(\log q(\la))\right]\right]~~, 
\end{equation}
where $\log q(\la)$ is the entropy of the fitness in the space of metabolic phenotypes and $D$ is a diffusion parameter
that, when fitting the model against experimental distribution \cite{de2016growth}, is seen to be of very low (adimensional) value: in the range $10^{-6}- 10^{-4}$.
It should be noticed that the term $\langle \lambda \rangle$ in principle renders the equation non-linear and its resolution challenging.
Given the low value of the diffusion parameter, in this paper we address a precise mathematical analysis of the small diffusion (i.e. $D$)
limit of the steady-state and time response of the
model. Quite remarkably, we will show  after  a rather non-standard WKB analysis that  the steady state and the time response of the model
are described by the ground state and the first excited state of a quantum particle
in a Airy potential plus a centrifugal term, i.e. that is governed by the Hamiltonian
\begin{equation*}
 H=-\frac{d^2}{dw^2}+w+\frac{a(a-2)}{4 w^2}
\end{equation*}
where {$w=D^{-\frac{1}{3}}(1-\la)$} and $a$ is the exponent of the vanishing of $q$ at the
maximum growth rate $\la_{max}$ allowed by the constraints.

Moreover, we will find in this regime scaling laws for growth fluctuations $\sigma$ and time response $\tau$ as a function
of the mean growth rate distance from the max $\la_{max}-\mla$:
\begin{align*}
& \sigma \sim \tau^{-1} \sim \la_{max}-\mla\sim D^{1/3} 
\end{align*}
suggesting that suboptimal (w.r.t to growth) {can have faster a response to perturbations}.

The paper is organised as follows. We will first outline in a background section the definition of the metabolic space and of the model.
Then we will report our analytical and numerical results for the steady state and its response.
We will finally {discuss the possible biological insights of our work} and draw out conclusions and perspectives.

\section*{Background}
The highly conserved set of enzymatic reactions in cells devoted to free energy transduction, the processing of small  organic compounds and the production of cell's building blocks is the so called {\em intermediate metabolism}. Nowadays this set can be reconstructed  directly from the genome by keeping track of the proteins encoded in the DNA that are devoted to metabolic functions\cite{schellenberger2010bigg}.
For time scales slower than diffusion ($\simeq10^{-3}$s in E.Coli\cite{milo2015cell}), metabolic  dynamics   can be modelled as a well-mixed chemical reaction  network  in which $M$ metabolites participate in $N$ reactions  with the  stoichiometry encoded in a matrix $\mathbf{S}=\{S_{\mu r}\}$.
Further, a phenomenological reaction consisting of a drain in fixed proportions  of biomass metabolites {is usually}  added in order to model cell growth.
The concentrations $c_\mu$ change in time according to mass-balance equations
\begin{equation}
\dot{\mathbf{c}} = \mathbf{S \cdot v}
\end{equation}
where $v_i$ is the flux of the reaction $i$ that is in turn a  (possibly experimentally unknown) function of the concentration levels  $v_i(\mathbf{c})$.
For timescales faster than gene expression ($\simeq 10^{2}$s\cite{milo2015cell}) we can consider the steady state (homeostasis) and bound the 
fluxes  in certain ranges $v_r \in [v_{r}^{{\rm min}},v_{r}^{{\rm max}}]$ that take into account thermodynamical irreversibility, kinetic limits and physiological constraints.
The set of constraints
\begin{eqnarray}\label{eq3}
\mathbf{S \cdot v}=0, \nonumber \\
v_r \in [v_{r}^{{\rm min}},v_{r}^{{\rm max}}]
\end{eqnarray}  
defines  a convex polytope in the space of reaction fluxes that we will call the space of  feasible metabolic phenotypes.
An uniform sampling of such space can be performed by an an hit-and-run Monte Carlo Markov chain\cite{Turcin:1971} in feasible
times\cite{Lovasz:1999p4121} after suited preprocessing\cite{de2015uniform}.
{The marginal growth rate distribution $q$ is in principle a
piece-wise polynomial function (see \cite{Hormann04}, Chapter 11.2), but previous samplings on different models show that it is globally}
well fitted by the formula 
\begin{equation}
\label{eq:Q definition}
q(\lambda) \propto \lambda^b(\lambda_{\max}-\lambda)^a~~ .
\end{equation} 
Here $a$ ({respec.} $b$)
stands approximately for $D-d-1$, where $D$ is the dimension of the polytope and $d$ the dimension of the
subspace where $\lambda$ is maximised ({resp..} minimised), i.e.  $a\simeq D-1$ for the case of a vertex
\footnote{{The difference between the theoretical and the fitted value of $a,b$ depends on the choice of the fitting function.
For example, in the model considered in \cite{de2015uniform} the theoretical value of $b$ is $4$ and the fitted value is $3.6$}}.
\begin{figure}[h!!!!!]\label{fig1}
\begin{center}
\includegraphics*[width=0.6\textwidth,angle=0]{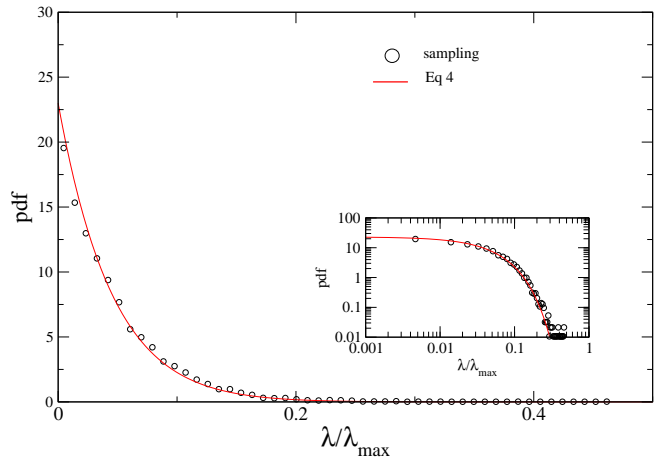}
\caption{Marginal growth rate probability distribution from an uniform sampling of the steady states of the catabolic core
of the genome scale model metabolic network E Coli iAF1260 with Monte Carlo methods in glucose limited minimal medium
(maximum uptake $10$mmol/GDWh, for a maximum growth rate of $0.874$h$^{-1}$),  compared with formula {(4)} with $b=0$, $a=22$.
{Inset: same plot in logarithmic scale.}}
\end{center}
\end{figure}
In the following we will consider the problem in the general form but, when comparing to numerical simulation 
we will  consider the E. Coli catabolic core from the  genome scale reconstruction iAF1260 in a glucose limited minimal
medium\cite{orth2011comprehensive}. 
After removing leaves, this is a chemical reaction network consisting of $N=86$ reactions among $M=68$ metabolites,
the dimensionality of the space of steady states is $D=23$. 
The flat distribution of growth rates in glucose limited minimal medium is well fitted by the parameters $a=22$, $b=0$ (Figure 1).

We will consider now the {phenomenological} dynamical model defined in\cite{de2016growth} that we briefly recall here.
It consists  of a standard exponential growth  coupled with diffusion in the aforementioned space.

\begin{figure}[h!!!!!]\label{fig1b}
\begin{center}
\includegraphics*[width=0.4\textwidth,angle=270]{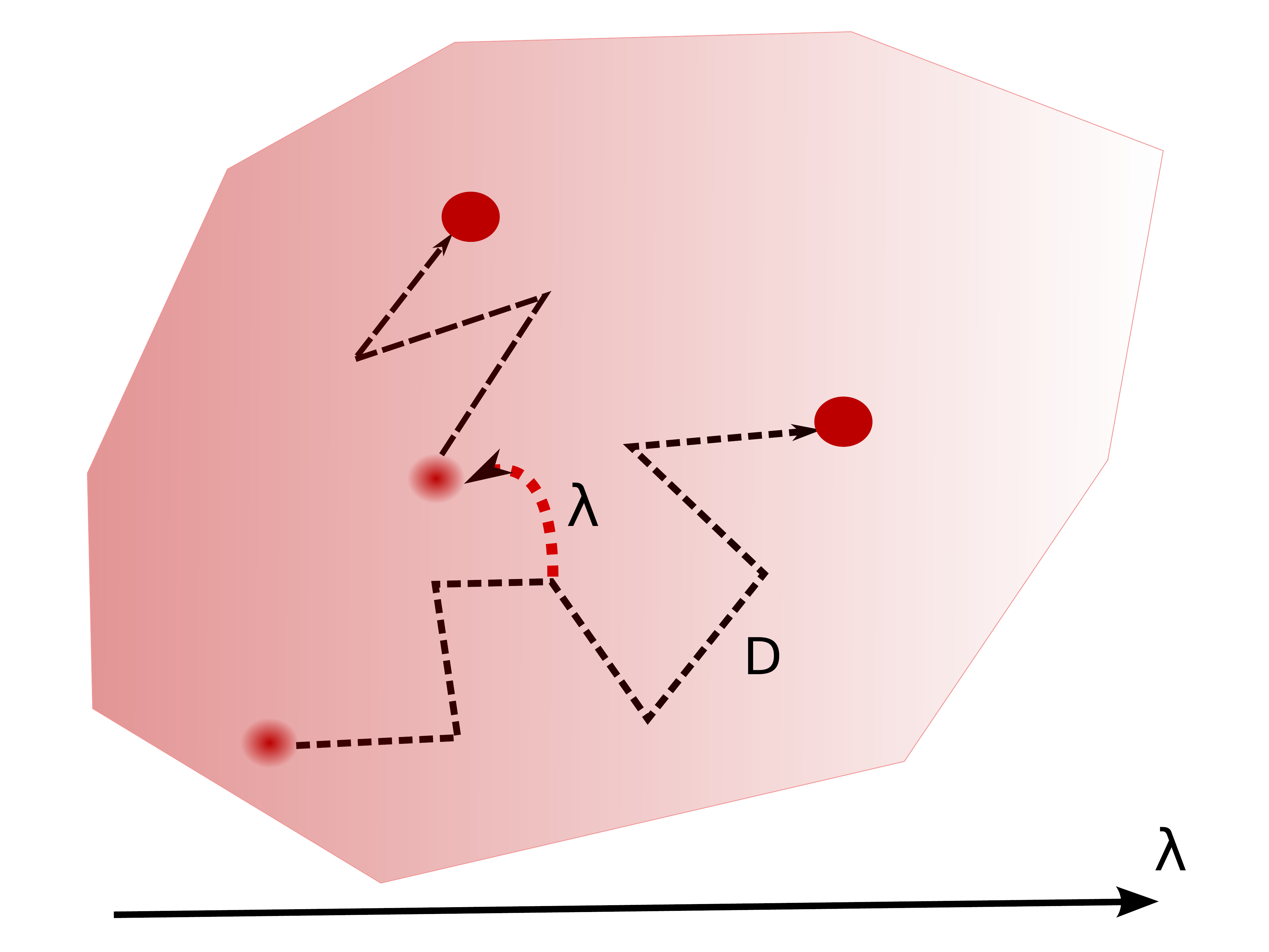}
\caption{Sketch of the dynamics of the model: random walkers move with a diffusion constant $D$ in a
convex space and duplicate with a rate $\lambda$ proportional to the distance along a certain direction.}
\end{center}
\end{figure}

If we denote by $N(\lambda)$ the number of bacteria growing at rate $\lambda$
this follows the equation 
\begin{equation}\label{eq:diff}
\dot{N}(\lambda)=\lambda N(\lambda) +\sum_{\lambda'} \left[W(\lambda' \to \lambda)N(\lambda')-W(\lambda \to \lambda')N(\lambda)\right] ~~,
\end{equation}
where $W(\lambda \to \lambda')$ stands for the transition rate from a phenotype with growth rate $\lambda$ to one with growth rate $\lambda'$. 
Metabolic changes are assumed to occur so that the space  of  phenotypes is explored in an unbiased way (i.e. we will impose the detailed balance condition with respect to the flat distribution $q$) and with a fixed time scale  $T_\epsilon$:
\begin{eqnarray}\label{eq:detailed}
W(\lambda' \to \lambda)q(\lambda')=W(\lambda \to \lambda')q(\lambda)~~, \\
\sum_{\lambda'} W(\lambda \to \lambda')=\frac{1}{T_\epsilon}~~. 
\end{eqnarray} 

If we consider a discrete random walk implementation of this dynamics, where only transitions $\lambda\to\lambda\pm\delta$  with sufficiently small
$\delta$ are allowed, 
in the limit $\delta,T_\epsilon \to 0$ ($\frac{\delta^2}{T_\epsilon} \to D$ constant), one obtains,
upon expanding, the non-linear Fokker-Planck equation for the probability
$p(\lambda)=\frac{N_\lambda}{\sum_{\lambda'}N_{\lambda'}}$:
\begin{equation}\label{fp}
\dot{p}(\lambda) = (\lambda-\langle \lambda \rangle) p(\lambda) +D \left[\frac{\partial^2 p}{\partial\lambda^2} -\frac{\partial}{\partial\lambda} \left[p(\lambda) \frac{\partial}{\partial\lambda}(\log q(\lambda))\right]\right]~~,
\end{equation}
where $D$ represents the ``diffusion constant'' of the population in the phenotypic space.   

For sake of simplicity we will re-scale $\lambda$ with $\lambda_{\max}$, that implies
$t \to t\lambda_{\max}$ and $D \to D/\lambda_{\max}^3$. Once the space (encoded by $q$) is fixed ,  $D$ is the only free parameter of this model. 
Previous  fits\cite{de2016growth} of experimental distributions of microbial growth rates with the stationary numerical solutions of this model
retrieve very low values for $D$ (adimensional), i.e. in the range $10^{-6}-10^{-4}$. This justifies the approach that we will present in the next section,
where we will perform a WKB approximation of the equation considering $D$ as small parameter. 
\begin{rem}
{Our choice of the detailed balance equation (\ref{eq:detailed}) with respect to the flat distribution $q$ corresponds to neglect growth control by cells. Such a control could be implemented,
at a phenomenological
level, upon adding a potential term in the non-replicator FK part of the
equation (\ref{fp}). This corresponds to a transformation of the kind  $q(\la)\to  q(\la) e^{v(\la)}$ for some
function $v(\la)$.
However, as we shall see in what follows, our results are unchanged 
by a transformation of this kind because they depend
exclusively on the order of vanishing of $q$ at $\la_{max}$}.
\end{rem}
\begin{rem}
 {The partial differential equation (\ref{fp}) is the continuous limit of the differential-difference equation (\ref{eq:diff}). This approximation
is valid only when considering  large populations $N\gg 1$. }
\end{rem}

\section*{Results}
In this section we will make a precise analysis of the steady state solution and its response to perturbation in the limit of slow diffusion using the WKB approximation that we will compare with numerical computations finding a good agreement, highlighting  scaling laws for growth fluctuations and time response. {The WKB anaysis of nonlinear PDEs is notoriously difficult (see e.g \cite{mara11}) and, as as consequence, the mathematical methods of this Section are rather non-standard.} 

\subsection*{Steady state solution}
Here we consider the steady state of the Fokker-Planck equation (\ref{fp}). For sake of definiteness we stick to a
marginal distribution $q$ of the form (\ref{eq:Q definition}), namely $q(\la)=\la^b(1-\la)^a$ for $a,b \in \bb{R}, a >1$.
However we notice that the mathematical theory is largely independent on the exact form of $q(\la)$.
In particular, as we will show below, the asymptotic form of the steady state,
in the limit $D \to 0$, depends only on the local behaviour of $q$ for $\la \to 1$.

\subsubsection*{A boundary value problem}
In this subsection we show that the nonlinear differential equation describing the steady state solution $p_s$
can be mapped into a boundary value problem for a linear differential equation - a simplification of great usefulness.

The equation for a stationary solution $p_s$ reads
\begin{equation}\label{eq:stationary}
 \frac{\partial^2 p}{\partial\lambda^2} -\frac{\partial}{\partial\lambda}
\left[p(\lambda) \frac{\partial}{\partial\lambda}(\log q(\lambda))\right]
 +\big( \frac{\lambda-\langle \lambda \rangle}{D}\big)p(\lambda)=0
\end{equation}
Here $\lambda \in [0,1]$. $a>1,b\geq 0$ are integer numbers, $D$ is a a positive real number and
$\mla=\frac{\int_0^1 \la p(\la) d \la}{\int_0^1 p(\la) d\la}$ for any non-negative function $p(\la)$.

As a first step we must discuss which boundary conditions the steady state solution satisfies.
The Fokker-Planck equation (\ref{fp}) is only well-defined if we impose the conservation of the total probability
$\int_0^1\frac{d}{dt}p(\la,t)$, which reads
\begin{equation}\label{eq:conservation}
 \int_0^1  \left[\frac{\partial^2 p}{\partial\lambda^2} -\frac{\partial}{\partial\lambda}
\left[p(\lambda) \frac{\partial}{\partial\lambda}(\log q(\lambda))\right]\right] d\la = 0
\end{equation}

We notice that the above condition allow us to transform the nonlinear differential equation (\ref{eq:stationary}) into a linear one
by substituting the nonlinear term $\mla$ with an arbitrary constant $\mu$.
Indeed, if $p_s$ satisfies the linear ODE 
\begin{equation}\label{eq:modstationary}
 \frac{\partial^2 p}{\partial\lambda^2} -\frac{\partial}{\partial\lambda}
\left[p(\lambda) \frac{\partial}{\partial\lambda}(\log q(\lambda))\right]
 +\big( \frac{\lambda-\mu}{D}\big)p(\lambda)=0 
\end{equation}
together with the condition (\ref{eq:conservation}) then $\mu=\mla$.
In fact, integrating (\ref{eq:modstationary}) and using (\ref{eq:conservation}) one gets
\begin{align*}
 \int_0^1 \big( \la - \mu )p(\la) d\la +D \int_0^1 \left[\frac{\partial^2 p}{\partial\lambda^2} -\frac{\partial}{\partial\lambda}
\left[p(\lambda) \frac{\partial}{\partial\lambda}(\log q(\lambda))\right]\right]~~= 0 \, \\ \Rightarrow 
\mu= \frac{\int_0^1 \la p(\la)}{\int_0^1p(\la) d \la} \; . 
\end{align*}

The linear ODE (\ref{eq:modstationary}) has two Fuchsian singularities at $\la=1,\la=0$.
This is more easily seen expanding the term with the logarithm and write the equation in the form
\begin{equation}\label{eq:esplicita}
 p''(\la)-\big(\frac{b}{\la}+\frac{a}{\la-1}\big)p'(\la) +
 \big( \frac{\lambda-\mu}{D} + \frac{b}{\la^2}+\frac{a}{(\la-1)^2}\big)p(\lambda)=0 
\end{equation}

Remarkably, we can use local analysis around the singularities
to map the conservation of probability into boundary conditions for $p_s$.

Close to $\lambda=0$ any solution will look
like $\lambda^{\rho}\big( 1 + O(\la) \big)$ \footnote{We remark that in the case $b=1$, one solutions behave as $\la$ and the other as $\la \log \la$.}
where $\rho$ satisfies the \textit{indicial equation}
$$
\rho(\rho-1)-b \rho+b=0 \quad \Longrightarrow \rho=1,b  \; .
$$

Similarly, close to $\lambda=1$, any solution will look like
$(\lambda-1)^{\rho}\big( 1 + O(1-\la) \big)$ where $\rho$ satisfies the \textit{indicial equation} equation
$$
\rho(\rho-1)-a \rho+a=0 \quad \Longrightarrow \rho=1,a \; .
$$
The conservation equation (\ref{eq:conservation}) is easily integrated to get
\begin{align*}
 \frac{\partial p}{\partial \la}(1,t)- \frac{\partial p}{\partial \la}{\la}(0,t) - \lim_{\epsilon \to 0} \big( p(1-\epsilon,t) (\frac{b}{1-\epsilon}+ \frac{a}{-\epsilon})
 -p(\epsilon,t) (\frac{b}{\epsilon}+ \frac{a}{\epsilon-1}) \big) =0
\end{align*}
Taking into account the possible local behavior at $\la=0,1$ of the solution $p$, the above conditions imply that
$p_s(\la)=(1-\la)^a$ for $\la \sim 1$ and - 
if $b\geq 1$ - that $p_s(\la)=\la^b$ for $\la \sim 0$. On the contrary, in the case $b=0$,
$p_s$ satisfies at $0$ the Robin-type boundary condition $p'_s(0)+a p_s(0)=0$

We have thus arrived to the a first characterization of the steady state.
\begin{lem}\label{lem:char1}
The steady state $p_s$ of the Fokker-Planck equation (\ref{fp}) is a non-negative solution of the following boundary value problem for the linear
ODE (\ref{eq:modstationary}):
\begin{itemize}
 \item[(i)] $p_s$ satisfies the linear ODE (\ref{eq:modstationary}) for some $\mu \in \bb{R}$. 
 \item[(ii)]Close to $\la=0$,  $p_s(\lambda) \propto \lambda^b $ if $b \geq 1$ and $p_\la(0)+a p(0)=0 $ if $b=0$.
 \item[(iii)]Close to $\la=1$, $p_s(\la) \propto (1-\la)^a$ 
\end{itemize}
Given a function $p$ satisfying (i,ii,iii) above then
\begin{equation*}\label{eq:mutomean}
 \mu= \mla \; .
\end{equation*}
\end{lem}

It is well-known \cite{codd55} that the above boundary value problem admits an infinite discrete set of (eigen-)solutions but
one and only one positive solution,
namely the ground state of the problem.
This will be discussed more thoroughly in the next paragraph.

\subsubsection*{Variational problem}

In this subsection we show that the steady state satisfies a variational problem. This will allow us to prove that the steady state is unique,
that the mean $\mla_s$ of the state state is a decreasing function of the diffusion parameter $D$ and, finally, that
for $D \to 0$ the mean value $\mla$ approaches $1$ with rate $D^\frac13$, namely $1-\mla_s= O(D^{\frac{1}{3}})$. 

In order to use the powerful variational techniques, as a first step we must transform the ODE (\ref{eq:modstationary})
into the standard Sturm-Liouville form.
This is very simple. We just have to define the new function
$t(\lambda)=p_s(\lambda)/q(\lambda)$ that satisfies the ODE
\begin{equation}\label{eq:tovariational}
 -D (q(\lambda) t'(\lambda))'-\lambda q(\lambda )t(\lambda)=-\mu q(\lambda) t(\lambda) \; .
\end{equation}
After our previous discussion $p_s(\la)$ satisfies the same boundary condition as $q(\la) $ and thus $t(\la)$
is a regular at $\la=0,1$.

As a second step, we  introduce the functionals
\begin{align}\label{eq:functionals}
 E[t]=\frac12\int_0^1  q(\lambda) \big(D t'(\lambda)^2 - \lambda t(\lambda)^2 ) d\lambda \, , \; 
 Q[t]=\frac12\int_0^1 q(\lambda) t(\lambda)^2 d\lambda
\end{align}
whose variational derivatives are easily computed as
\begin{align*}
  \frac{\delta E[t] }{\delta t(\lambda)}=-D (q(\lambda) t'(\lambda))'-\lambda q(\lambda )t(\lambda) \, , \;
   \frac{\delta Q[t] }{\delta t(\lambda)}= q(\lambda) t(\lambda)
\end{align*}
Using the principle of the Lagrange multipliers, we deduce that the differential equation (\ref{eq:tovariational}) for the function $t$
is just the equation for a critical point of the functional $E[t]$ subject to the
constrain $Q[t]=1$. In this interpretation $\mu$ is minus the Lagrange multiplier.
Since $t(\la)$ is by definition positive, the standard
Sturm-Liouville theory -see e.g. \cite{walter96}- asserts that $t$ is the unique minimum of the functional.

We have now arrived to a second - variational - characterization of the steady state $p_s$.
\begin{lem}\label{lem:char2}
The following statements
are equivalent:
\begin{enumerate}
 \item $\mu$ is such that there exists a positive solution of the boundary value problem for the liner ODE (\ref{eq:modstationary}).
 \item  $t(\la)=\frac{p(\la)}{q(\la)}$ is the unique minimum point, in the space $H^1([0,1])$ \footnote{$H^1([0,1])$ is simply the space of functions
 such that $\int_0^1t'(\la)^2 d\la =1$. In case $b=0$ the Robin boundary condition $p'(0)+ap(0)=0$ must be imposed.},
 of the functional $E[t]$ subject to the constrain $Q[t]=1$. In a formula
 \begin{align*}
\mu=- \inf_{Q[t]=1} E[t] 
\end{align*}
\item Equivalently, $\mu$ is minus the minimum of the Rayleigh quotient $\frac{E[t]}{Q[t]}$:
\begin{equation}\label{eq:freeminimum}
 \mu=-\inf_{t \neq 0}\frac{E[t]}{Q[t]}=\sup_{t \neq 0}\frac{-E[t]}{Q[t]}
\end{equation}
\end{enumerate}
\end{lem}

As we anticipated using the above variational characterization we can infer some useful information about the
steady state $p_s$ and its mean. For convenience of the reader and for future reference, we collect this
information in the following Theorem.
\begin{thm}\label{lem:var2}
 \begin{itemize}
  \item[(i)] Fixed $a,b$ and $D$, there exists a unique normalized stationary solution $p_s(\la)$ of the FP equation.
  \item[(ii)] Fixed $a$ and $b$, the mean $\mla_s(D)$ of the unique stationary solution $p_s$ is
  a monotone decreasing function of $D$.
  \item[(iii)] Fixed $a$ and $b$, we have
  \begin{equation}\label{eq:1-lambda}
   1-\mla_s(D)=O(D^\frac{1}{3})
  \end{equation}
and in particular
\begin{equation*}
 \lim_{D \to 0}\langle \lambda \rangle_s(D)=1
\end{equation*}

 \end{itemize}

\end{thm}
We analyze the three statements separately. Statement (i) is equivalent to the uniqueness of the ground state for Sturm-Liouville variational
problem; this is a standard results, see e.g. \cite{codd55} Chapter 9 and \cite{walter96} Chapter 6.
Statement (ii) is a direct consequence of the fact that
the functional $E[t]$ is a monotone decreasing function of $D$ while the constraint $Q=1$ is independent of $D$.
Finally statement [(iii)] is obtained by minimizing the functional $E[t]$ on a family of Gaussian test functions whose mean and standard deviation
depend on $D$.  These computations can be found in the Appendix below.

\begin{rem}
 We notice here that the 
exact minimizer in the small $D$ limit is actually not a Gaussian function but a different function (\ref{eq:at1}),
related to the Airy function that we will introduce in the next paragraph.
However the Gaussian approximation reproduces correctly the exponent of the asymptotic behavior, namely $D^{\frac13}$, of both the mean $1-\langle \lambda \rangle$
and of the standard deviation $\sigma $of the stationary distribution.
\end{rem}
In Figure 2 (left) we show a comparison of this predicted scaling of the average growth rate (as well as of its standard deviation,
Figure 2 right, see next section) with numerical simulations for the case of the core catabolic core of E.Coli  where we find a very good agreement for
low values of $D$.
\begin{figure}[h!!!!!]\label{fig2}
\begin{center}
\includegraphics*[width=0.45\textwidth,angle=0]{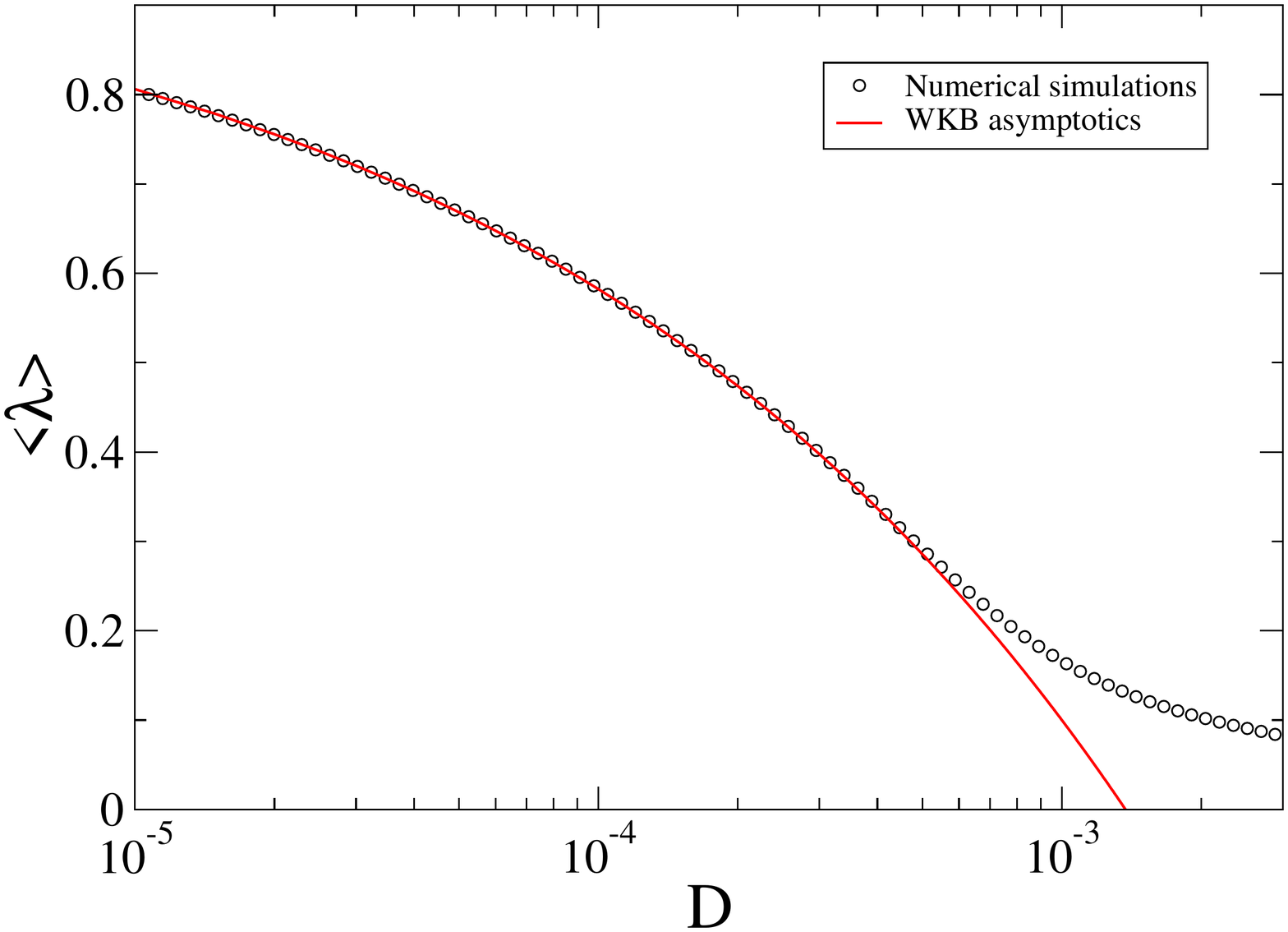}
\includegraphics*[width=0.45\textwidth,angle=0]{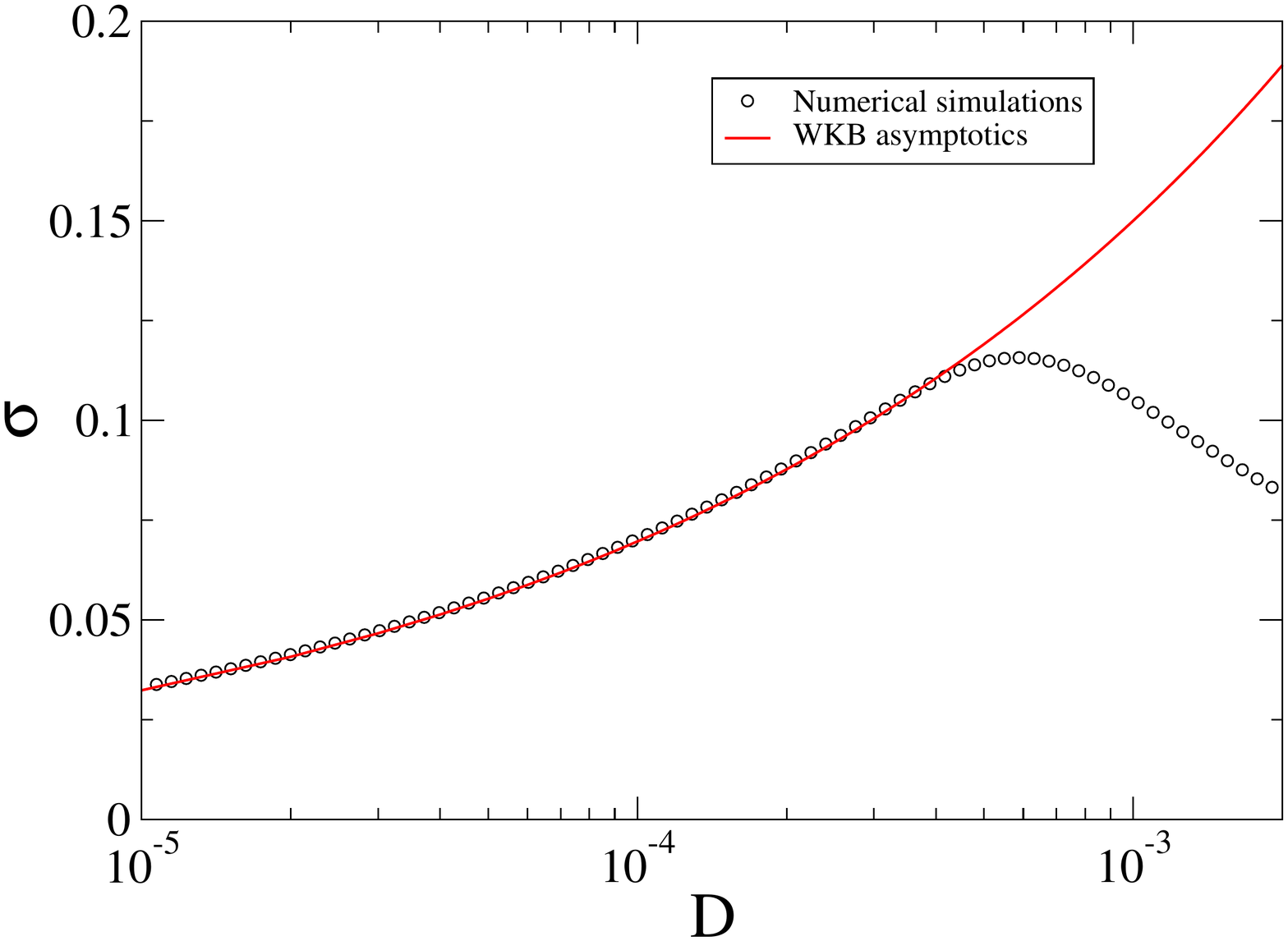}
\caption{Left: Average growth rate $\langle \la \rangle$ as a function of $D$ from numerical simulations compared with the
asymptotic WKB formula $\langle \la \rangle \simeq 1-A D^{1/3}$.
Right: Growth rate standard deviation $\sigma$ as a function of $D$ from numerical simulations compared with the
formula WKB asymptotic scaling $\sigma\propto D^{1/3}$.
Computation are done for a model mimicking the core catabolism of E Coli $a=22, b=0$. {In both cases simulations are compared directly  with WKB asymptotics without fitting parameters}}
\end{center}
\end{figure}

\subsubsection*{WKB approximation and the exact asymptotic expansions}
In this Section we discuss the asymptotic analysis of the steady state distribution in the \textit{WKB} limit $D \to 0$.
For sake of simplicity we present here the main results, together with some heuristic justification, and we leave
the details and the full mathematical justification to the Appendix.

As we have shown using the variational method, the mean $\mla_s$ of the steady state $p_s$ approaches $1$ with rate at least $D^{\frac13}$,
that is $\mla_s=1-A D^{\frac13}+o(D^{\frac13})$ ($A \geq 0)$.
Since the distribution takes value from $[0,1]$, we deduce that it is \textit{concentrated} in an interval of amplitude $D^{-\frac13}$
on the left side of $\la=1$.
In order to analyze the steady state in this interval we rescale the variable $w(\la)=D^{-\frac{1}{3}}(1-\la)$ and the steady-state
\begin{equation}\label{eq:at1}
 p_s(\la)=D^{-\frac{1}{3}} w^{\frac{a}{2}}Y_a(w) \big( 1 + O(D^{\frac12})  \big)\; .
\end{equation}
From the differential equation (\ref{eq:esplicita}) satisfied by $p_s$, one easily computes that the rescaled distribution $Y_a(w)$
satisfies
the Airy-like differential equation
\begin{equation}\label{eq:diffmodel}
 y''(w)=\big( w-A + \frac{a(a-2)}{4 w^2} \big) y(w) \;, \qquad w\geq 0 \; .
\end{equation}
More precisely, as we show in the Appendix, $Y_a(w)$ is unique solution of the above differential equation
satisfying the requirements
\begin{itemize}
 \item[(i)]$Y_a(w)$ is non-negative. More precisely $Y_a(w)>0$ for $0<w<+\infty$.
 \item[(ii)]$Y_a(w) \propto w^{\frac{a}{2}}\big( 1 + O (w) \big)$ for $w \to 0$.
 \item[(iii)]$\lim_{w \to \infty}Y_a(w)=0$.
 \item[(iv)]Normalization: $\int_0^{\infty}w^{\frac a2}Y_a(w) dw =1$.
\end{itemize}
Properties (i-iv) above reflect the corresponding properties of the steady-state, respectively $p_s$ is positive, $p_s \propto (1-\la)^a$, $p_s$
is localized close to $\la=1$ and $p_s$ is normalized.
Because of properties $(i,ii,iii)$ above, the number $A$ is actually the ground state of the Hamiltonian 
\begin{equation}\label{eq:AiryHamiltonian}
 H=-\frac{d^2}{dw^2}+ w +\frac{a(a-2)}{4 w^2} 	\;.
\end{equation}
and $Y_a(w)$ is the corresponding normalized eigenfunction. Since the potential is positive then $A$ is a strictly positive number.

For large $a$, a standard quantum-mechanical estimates based on the minimization of the potential yields
$A \sim \frac{3\big( (a(a-2)) \big)^{\frac23}}{16^{\frac13}}$.
We notice that the full spectrum of the Hamiltonian (\ref{eq:AiryHamiltonian}) is actually known to be computable via the Bethe Ansatz equation
of the Quantum-KdV equation (or the conformal limit of the Six Vertex Model), see \cite{dorey07ode,masoero15}.
However no explicit solution by means of special functions exists unless $a=2$, in which case the spectrum of
(\ref{eq:AiryHamiltonian}) coincide with minus the zeros of $Ai'$ (e.g. $A(a=2)=1.0188...$). In Fig. 3 (top)
we show the form of the function $w^{a/2}Y_a(w)$ computed by expanding in Airy function upto the $6^{th}$ order for the case $a=22$ as well as
(Figure 3, bottom) the stationary distributions that can be obtained from it upon rescaling  $w(\la)=D^{-\frac{1}{3}}(1-\la)$ compared with numerical simulations, where we find a very good agreement.  

\begin{figure}[h!!!!!]\label{fig3}
\begin{center}
\includegraphics*[width=0.6\textwidth,angle=0]{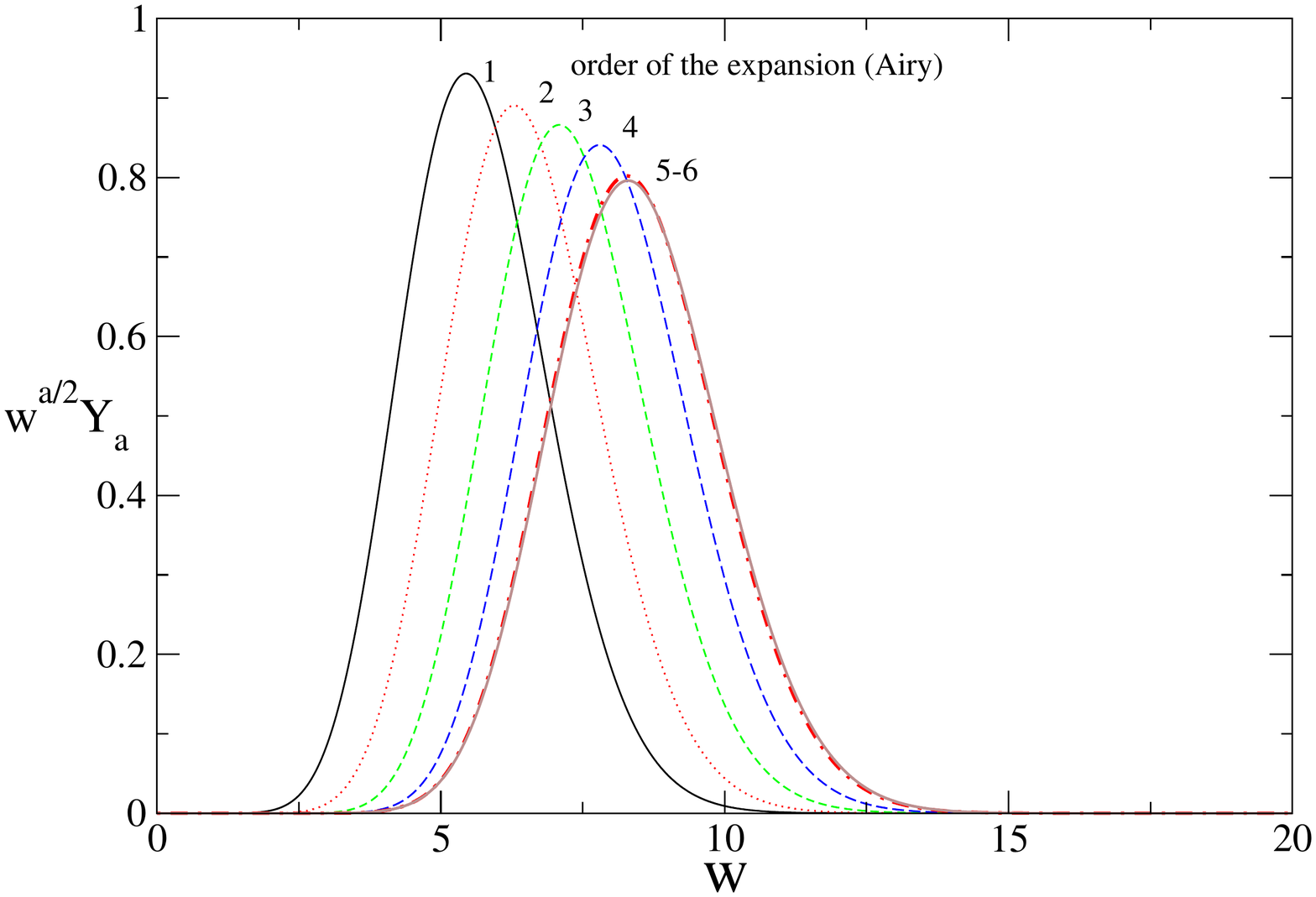}
\includegraphics*[width=0.7\textwidth,angle=0]{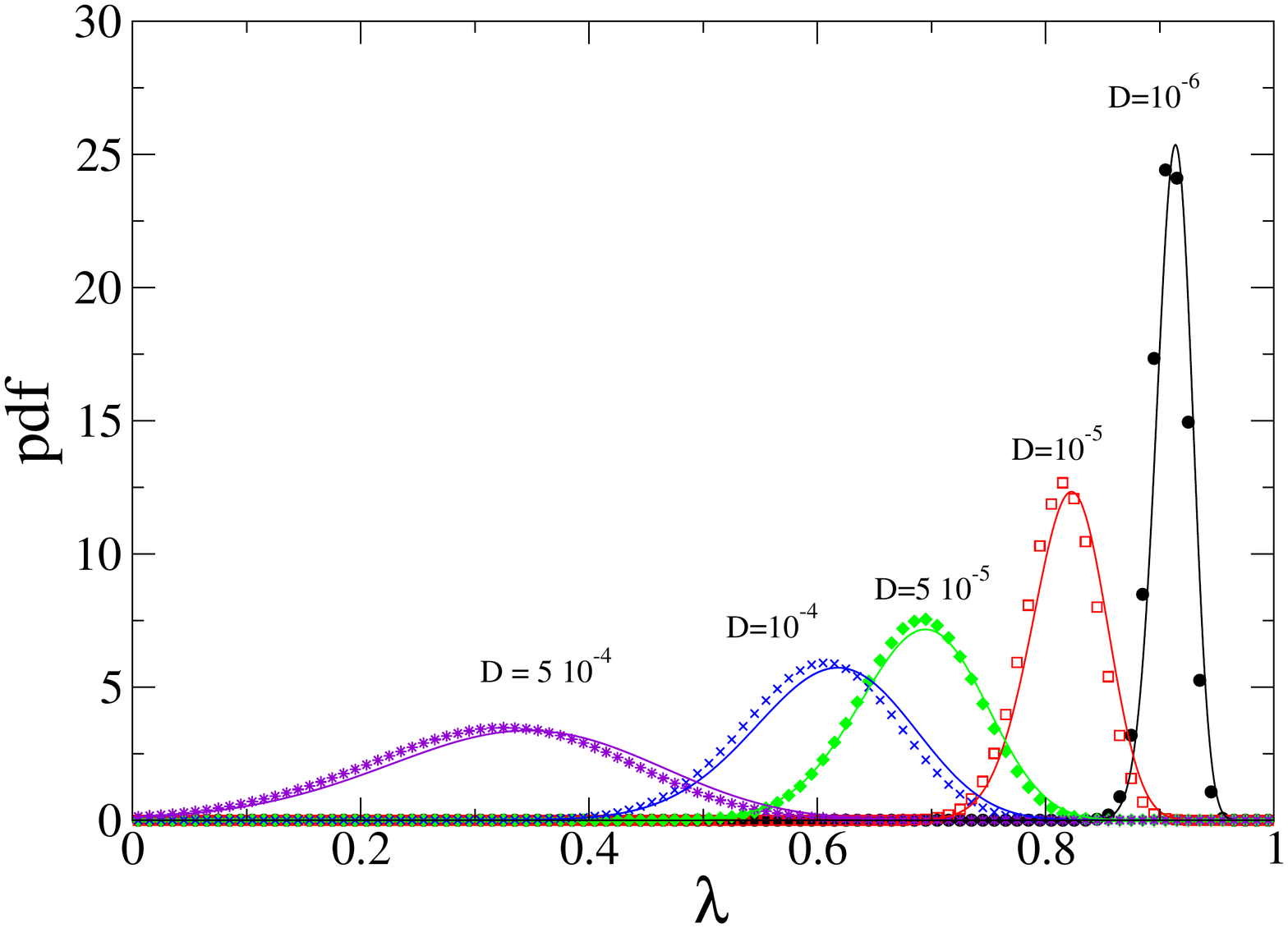}
\caption{Top: Normalized ground state function of the Airy-like Hamiltonian (), $w^{a/2}Y_a(w)$ obtained with an expansion in Airy functions upto to the $6^{th}$. Bottom: Steady state probability distributions of the (normalized) growth rates for several values of $D$ for numerical simulations (symbols) and WKB approximation (lines), with $a=22$, $b=0$. {In all cases simulations are compared directly with WKB asymptotics without fitting parameters.}}
\end{center}
\end{figure}

Apart from the mean $\mla$, all higher momenta of the steady state can be computed from $Y_a(w)$, see equation
(\ref{eq:highermomenta}) in the Appendix. The first two  momenta are particularly important
for data analysis. 
Letting $A$ be, as above,the ground state energy of the Airy Hamiltonian (\ref{eq:AiryHamiltonian}) and $\Sigma$ be defined 
by the formula
\begin{equation}\label{eq:Sigma}
 \Sigma= \sqrt{\int_0^{\infty} (w-A)^2 w^{\frac a2} Y_a(w) dw} \; .
\end{equation}
we obtain the following asymptotic laws for the mean and standard deviation
\begin{align*}
& \mla_s=1-AD^{\frac13}+ O(D^{\frac12}) \\
&  \sigma_s= \Sigma D^{\frac13}+ O(D^{\frac12})
\end{align*}
As a consequence, eliminating the parameter $D$, we find that the following scaling law holds in the limit $\mla_s \to 1$
\begin{equation}\label{eq:scale1}
\sigma_s=\frac{\Sigma}{A} (1 -\mla_s) \mbox{ as } \la \to 1 \; .
\end{equation}


\subsubsection*{Relaxation time}
Here we  compute the relaxation time in the $D \to 0$ regime. We use the usual method of linear stability:
we linearize the FP equation around the steady state and
look for eigensolutions of the linearized operator. Namely, we add a small perturbation to the steady state
$p(\la)=p_0(\la) +\delta(\la)$ (with $\int_0^1 \delta(\la) d\la=0$ to ensure that $\int_0^1 p(\la) d\la=1$)
and look for the smallest positive number $\eta$ such that
the perturbation relaxes as $\delta(x,t)=e^{-\eta t}\delta(x,0)$. By definition, the relaxation time $\tau$ is then simply $\eta^{-1}$.

The linearized equation is as follows
\begin{equation*}
 - \eta \delta(\la)= (\la -\mla_s) \delta - p_s(\la) \int_0^1 \la \delta(\la) d \la +
D \left[\frac{\partial^2 p}{\partial\lambda^2} -\frac{\partial}{\partial\lambda} \left[p(\lambda)
\frac{\partial}{\partial\lambda}(\log q(\lambda))\right]\right] \; ,
\end{equation*}
where $\mla_s=\int_0^1 \la p_s(\la) d\la$.

In the Appendix, we show that $\eta$ has the following asymptotic expansion
\begin{equation*}
\eta=(\lambda^{(1)}-A) D^{\frac13} +o(D^{\frac13}) 
\end{equation*}
where $A$ is the ground-state energy of the Airy like Hamiltonian $H$ (\ref{eq:AiryHamiltonian}) and $\lambda^{(1)}$
the energy of the first excitation.

We therefore immediately deduce that the relaxation time scales as $D^{-\frac 13}$
\begin{equation*}
 \tau \sim  D^{-\frac 13}  \, .
\end{equation*}
and thus
\begin{equation}\label{eq:tauasymptotic}
 \tau \sim  \frac{1}{1-\mla_s}  \, \mbox{ as  } \mla_s \to 1 .
\end{equation}

{
\subsubsection*{Universality}
For sake of definiteness we have worked with the marginal distribution $q=\la^b(1-\la)^a$. However,
the asymptotic analysis is largely independent on the exact form of the marginal distribution. Indeed, the asymptotic results
depend exclusively on the order of vanishing of $q$ for $\la \to 1$.

Let in fact
consider the expansion of a general marginal distribution around $a=1$, $q(\la)=(1-\la)^a\big(c_0+c_1 (1-\la)+ O((1-\la)^2) \big)$.
We introduce as above the rescaled distribution
\begin{equation*}
 p_s(\la)=D^{-\frac{1}{3}} w^{\frac{a}{2}}Y_a(w) \big( 1 + O(D^{\frac12})  \big)\; .
\end{equation*}
From the differential equation (\ref{eq:modstationary}) satisfied by $p_s$, one computes that the rescaled distribution $Y_a(w)$
satisfies the same Airy-like differential equation
\begin{equation*}
 y''(w)=\big( w-A + \frac{a(a-2)}{4 w^2} \big) y(w) \;, \qquad w\geq 0 \; ,
\end{equation*}
Therefore a change of the form of the marginal distribution, that does not alter the exponent $a$, does not change the leading order of the
asymptotic expansion.
Moreover, even when the exponent is altered, the scaling laws (\ref{eq:scale1},\ref{eq:tauasymptotic}) remain valid but for a modification of
the prefactor. }

\begin{figure}[h!!!!!]\label{fig2c}
\begin{center}
\includegraphics*[width=0.65\textwidth,angle=0]{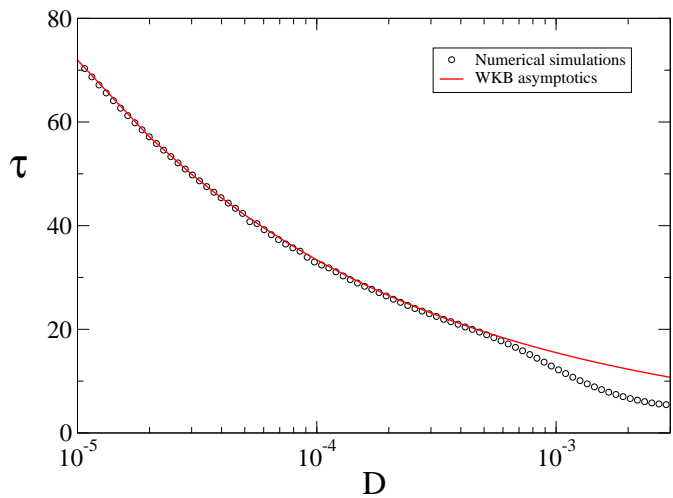}
\caption{Growth rate relaxation time $\tau$ as a function of $D$ from numerical simulations compared with the
formula WKB asymptotic scaling $\tau\propto D^{-1/3}$ for a model mimicking the core catabolism of E. Coli $a=22, b=0$. {Simulations are compared directly against WKB asympotic without fitting parameters} }
\end{center}
\end{figure}

{
\section*{Discussion}
The simple model we studied couples the standard flux-balance analysis with a diffusion in the space of phenotypes.
We discovered that this model predict two interesting scaling laws that are possibly significant to biology:
\begin{equation}\label{eq:scaling}
\lambda_s = \la_{max}-C_1\sigma \qquad
\tau = C_2/\sigma \; ,
\end{equation} 
if $\frac{\sigma}{\la_{max}} \ll 1 $. Here $\la_{max}$ is the maximal growth-rate allowed by the flux-balance analysis, and 
$\la_s$, $\sigma$ and $\tau$ are respectively the mean growth-rate, the standard deviation and the response time of the equilibrium distribution
of growth-rates. The scaling parameters $C_1,C_2$ depends exclusively on the a single parameter describing the flux polytope close to the
vertex (or face) maximizing the growth-rate.

The experimental data available in the literature do not span an  range wide enough  
in order to test unambiguosly the validity of the 
 the scaling laws (\ref{eq:scaling}), in particular in the regime of small fluctuations and high autocorrelation times.
However we do point out that the scaling is
in agreement with two emerging stylized facts of growth rate statistical dynamics of E. Coli, i.e. 
\begin{itemize}
\item The Fano factor is approximately constant across mildly varying conditions\cite{iyer2014scaling} 
\begin{equation}
F = \frac{\sigma_\lambda}{\langle\lambda\rangle} \simeq \textrm{const.}
\end{equation}
\item Mother-daughter correlations (i.e. one generation auto-correlation) are approximately constant across mildly varying conditions\cite{taheri2015cell}.
Assuming that the auto-correlation has actually an exponential decay, and since the decay of the auto-correlation is governed by the relaxation
time, then we can estimate the relaxation time $\tau$
\begin{equation}
C_{MD} \simeq e^{-\frac{1}{\tau \langle\lambda\rangle}} \simeq \textrm{const.} 
\end{equation} 
\end{itemize}
From these experimental stylized facts it follows a scaling of the form
\begin{equation}
\tau \simeq  \frac{C_2}{\sigma} \mbox{  where  }
C_2=\frac{F}{\log(\frac{1}{C_{MD}})}
\end{equation}
The pre-factor computed from the experiment \cite{taheri2015cell},
where $F\simeq 0.15$ and $C_{MD} \simeq 0.4$, is $C_2\simeq 0.4$. A standard choice for the polytope of fluxes for E. Coli, with the parameter
$a=22$, yields
$C_2\simeq 2.4$.  
Such  a discrepancy in the pre-factor either hints at the fact
that fewer degrees of freedom are controlling the growth fluctuations, or suggests that 
our model, as it is typical for coarse-grained statistical mechanics models,
predict correctly the scaling laws but not the precise pre-factors.}

{
Further, the scaling laws we have found show that
suboptimal phenotypes could have a faster adaptation. They can therefore be more favourable, as it can be seen analysing a simple linear model
for up-shift response.

Suppose that the system is stationary and at time $t=0$ is perturbed in such a way that the diffusion parameter of the Fokker-Planck (\ref{fp}) equation
is unaltered.
Assuming a linear response, the mean growth rate -as a function of time- has the following form
\begin{equation}\label{eq:menawithtime}
\mla (t)=\lambda_s+ \Delta \la e^{-t/\tau} \, .
\end{equation}
and its time-averaging on a time scale $\Delta t$ is
\begin{equation}\label{eq:averagedmean}
\overline{\lambda} = \lambda_s+ \frac{\Delta \la}{\Delta t} \tau \big( 1 -e^{-\Delta t/\tau} \big) \; .
\end{equation}
We can consider such a simple equation to model two different situations.
The first is when the system undergoes a stress (e.g. an antibiotic administration) at time $t=0$ and $\Delta t$
represents the time difference between two of such stresses. The second situation is an up-shift  (e.g. higher glucose
concentration) of duration $\Delta t$, so that the maximal growth $\la_{max}$ allowed by the external conditions in the time interval $[0,\Delta t]$
is bigger than at time $0$. In both cases the mean growth increases with time, therefore they are modelled by the formula (\ref{eq:menawithtime})
with a negative $\Delta \la$. 

Upon implementing  the scaling laws (\ref{eq:scaling}), there is a trade-off between increasing $\lambda_s$ and decreasing
$\tau$. This is controlled by a single parameter that we can take as the fluctuations $\sigma$.
We find that the function $\overline{\lambda}(\sigma)$, as soon as $\Delta t$ is not too small (the response time is in no case istantaneous),
is  maximised at a finite value of $\sigma>0$. 
This is pictorially shown in the Figure  6 below.
}

\begin{figure}[h!!!!!]\label{pep}
\begin{center}
\includegraphics*[width=0.65\textwidth,angle=0]{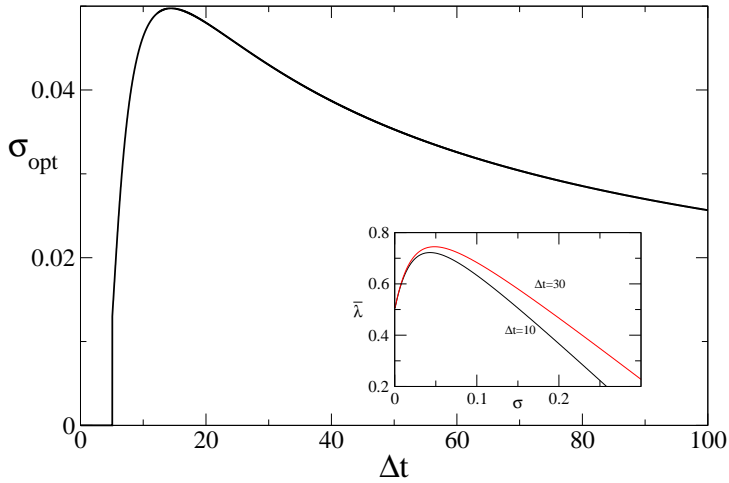}
\caption{The standard deviation $\sigma_{opt}$ maximizing  the averaged mean growth
$\overline{\la}$ as a function of the time-scale $\Delta t$.
for parameters  value $C_1=3$, $C_2=0.4$, $\Delta \la=-0.5$ for
equations (\ref{eq:scaling},\ref{eq:averagedmean}). Inset:  $\overline{\la}$ as a function of $\sigma$ for $\Delta t=10,30$}
\end{center}
\end{figure}

\section*{Conclusions}
The present richness of data in molecular biology could gain many insights from a systemic perspective.
This in turn could gain theoretical insights from physics, in particular from statistical mechanics, in terms of rigorous methods and quantitative modelling.
Many of the current analysis coupling cell metabolism and growth relies on the assumption of growth maximisation (FBA) that by definition cannot
assess fluctuations.

In this draft we studied  the {phenomenological} model defined in \cite{de2016growth} where the metabolic dynamics is connected to
cell growth through a simple diffusion-replication process in the space of metabolic phenotypes as it can be obtained from realistic
constraints-based modelling {within the approximation that cells lack control on their growth rate and in the limit of large population size}.
We have studied the small diffusion limit and shown that the steady states of the resulting non-linear Fokker Planck equation can be mapped through the WKB approximation
onto the ground states of a quantum particle in an Airy potential plus a centrifugal term, finding a good agreement with numerical simulations for
a case of interest, i.e a core catabolic network of E.Coli. {Remarkably the small diffusion asymptotics that we analysed,
depends just on one parameter describing the space of metabolic phenotypes
close to the face that maximizes the growth.}

In the asymptotic regime we have found scaling laws for growth fluctuations and time response that can be tested against experimental data.
We have found that increasing fluctuations lower the response time, and in turn fluctuations are, in this context akin to constrained
optimisation, proportional to the distance from the maximum growth rate. 
Apart from the experimental evidence of fluctuations, a strict maximisation of the growth rate shows a divergent response time even in this
simple diffusive framework. {As a consequence, in our model, sub-optimal mean values of the growth rate lead to faster adaptive response. We briefly discussed the case of a linear up-shift response in which, because of non-equilibrium dynamics, sub-optimal mean values eventually leads to
a higher growth rate}.

We finally outline  some interesting future perspectives that can stem from this work.
{Even though the data available in the literature do not yet allow a direct test of the scaling laws, the
fluctuation-response relation we have found is in agreement with two stylised facts, i.e. the constancy of the Fano
factor and of mother-daughter correlations}. Dedicated experiments are needed in
order to  prove the scaling laws that we have found, {in particular 
pointing out the possibility of emerging fluctuation response relations in biological networks, as it has been shown for the signalling network underlying bacterial chemiotaxis\cite{park2010interdependence}}. 
Moreover, since the model in principle  retrieves  dynamical trajectories for the entire metabolic network, one can compare against experimental data
the steady state distribution and time response of the all enzymatic fluxes predicted by the model.
{Comparison with experimental data will also allow a clearer interpretation and/or a direct estimate of the diffusion parameter of the model. Finally, at the mathematical level, it would be interesting to extend the model in order to allow cells control on their growth rate.
The control can be inserted in the model by modifying the flat measure of the space, but such a modification only affects
sub-leading terms in the WKB asymptotics
which are beyond the scope of the present paper.}

\section*{Appendix}
\subsection*{Best Gaussian Approximation}
In this section we prove Theorem (\ref{lem:var2}) (iii). We show using a variational estimate that, fixed the parameters $a$ and $b$,
the mean value $\mla_s$ of the steady state has the following asymptotic behaviour for $D$ small
\begin{equation*}
 \mla_s=1-O(D^{\frac 13})  \;.
\end{equation*}

To this aim we use the characterisation via the Rayleigh quotient (\ref{eq:freeminimum}), i.e.
\begin{equation*}
 \mla_s=\sup_{t \neq 0}\frac{-E[t]}{Q[t]}
\end{equation*}
where the functionals $E,Q$ are defined in (\ref{eq:functionals}).
We evaluate the Rayleigh quotient on a family of Gaussian approximation and find that the supremum on this family is $1-\kappa D^{\frac 13}+o (D^{\frac 13})$
for some unimportant $\kappa >0$.
Since the actual supremum is, by definition, a value between this estimate and $1$ (because $\mla<1$ for any function) then we conclude that
$\mla_s=1-O(D^{\frac 13})$, that is $\mla_s$ approaches $1$ with a rate \textit{at least} $\frac13$.

If $D=0$ the steady state solution is not well-defined, however the functionals still are. This reflects in the fact
while the Rayleigh quotient has a supremum, it has no maximum if $D=0$.
It is both instructive and useful to compute that the supremum for $D=0$ is $1$. This is easily seen
by evaluating the functionals on a family of approximations of a $\delta$-function centred at $\lambda=1$.

Let us thus define $G(x,\sigma)$ ($x,\sigma >0$) as the (positive) square root of the Gaussian distribution
with mean $1-x$ and standard deviation $\sigma$, namely
\begin{equation}\label{eq:gaussian}
 G(x,\sigma)=\big(2 \sigma \pi\big)^{-\frac12}e^{-\frac{(\la-1+a)^2}{\sigma^2}} \; .
\end{equation}

The Rayleigh quotient for these Gaussian approximation reads
$$-\frac{H[t]_{D=0}}{Q[t]}=1-\frac{\int_0^1 G^2 (1-\la) q(\la) d\la}{\int_0^1G^2 q(\la) d \la} \; . $$
This can be easily estimated with the help of the Laplace's method.
In fact, in the small $\sigma$ and $x$ limit the following formula holds  (recall that $q(\la)=\la^b(1-\la)^a)$)
\begin{align}\label{eq:laplaces}
 & \int_0^1 G^2(a,\sigma) q(\la) (1- \la)^k d\la=c_k x^{a+k} \big( 1 + O(x) \big)+ \\ \nonumber
& d_k \frac{\sigma^2}{2} x^{a+k-2}
  \big( 1 + O(x) \big) + 
  O(\sigma^3) ,  k\geq 0
\end{align}
for some unimportant constants $c_k>0,d_k $.
The above computations gives that the small $\sigma$ and $x$ limit of the Rayleigh quotient is
$1-\frac{c_1}{c_0} x \big( 1 +O(x) \big)+ O(\sigma)$. Therefore the supremum with respect to the Gaussian family -
attained by taking the limit $\sigma,x\to 0$ - is $1$. This proves that the supremum is actually $1$.

In order to estimate the supremum of the Rayleigh quotient for $D$ small but different from zero, we
can use the same family of test functions. However,
since for $D \neq 0$ the actual minimizer $t(\la)$ has a finite standard deviation
and the mean is smaller $1$, we let the parameters $x,\sigma$ of the test function $G$ to depend on the small
parameter $D$ and look for the best approximation possible.

More precisely, we first let $x = D^{\alpha}, \sigma =
D^{\beta}$ for some $\alpha,\beta >0 $ and obtain the estimate $\frac{-H[t]}{Q[t]} = 1- O \big(  D^{\gamma} \big)$ for some $\gamma$
depending on $\alpha,\beta$. Subsequently we optimise $\alpha,\beta$ in order to get the biggest possible $\gamma$. As it will turn out,
the optimal choice is $\alpha=\beta=\frac{1}{3}$ that yields $\gamma=\frac13$.

Assuming $\alpha \leq \beta$, after formula (\ref{eq:laplaces}) 
$ \int_0^1 G(x,\sigma)^2 q(\la) (1- \la)^k d\la \propto  D^{\alpha(a +k)} $.
Noticing that $G'(x,\sigma)^2= \frac{(1-\la+x)^2}{4 \sigma^4}G^2$, we can compute
the Rayleigh quotient in the small $D$ (and thus small $x,\sigma$) limit using the above formula.
Indeed, provided $\alpha \leq \beta$, we get
$$
1+\frac{H[t]}{Q[t]} = k_1 D^{1+2 \alpha-4\beta}+ k_2 D^{\alpha} + \mbox{ higher order terms}
$$
for some unimportant constants $k_1,k_2>0$.
The optimal choice of the parameters is given by the detailed balance conditions
$1+2\alpha-4\beta=\alpha$, $\alpha=\beta$ (thus $\alpha=\frac13$), which yields
the desired results $1+\frac{H[t]}{Q[t]} = O \big(D^{\frac 13} \big)$.

\subsection*{WKB Analysis}

Here we compute dominant asymptotics of the steady state $p_s(\la), \la \in[0,1]$ in the small $D$ limit, uniformly in the whole unit interval.
As we have shown in the main body of the paper, $p_s$ is a positive solution of the ODE
\begin{equation}\label{eq:modstationary2}
 p''(\la)-\big(\frac{b}{\la}+\frac{a}{\la-1}\big)p'(\la) +
 \big( \frac{\lambda-\mu}{D} + \frac{b}{\la^2}+\frac{a}{(\la-1)^2}\big)p(\lambda)=0 
\end{equation}
that close to the singularities $\la=0,1$ behaves as
\begin{equation*}
 p_s(\la) \propto \la^b , p_s(\la) \propto (1 -\la)^a .
\end{equation*}

The limit $D \to 0$ is a singular limit of the above ODE. In this limit there are in principle
three regions in which $p_s$ assumes different asymptotic behaviours \cite{fedoryuk93}. The first region is when $\la$ is close
to a singularity, namely $\la \sim 0,1$: in this region the solutions have a Bessel asymptotics.
The second region is the bulk, when $\la$ is away from the singularities
and $\mu$, i.e. $\la \neq0,1,\mu$: in the bulk the solutions have a WKB asymptotics. The third and final region is when $\la$ is close
to $\mu$ ($\mu$ is called the \textit{turning point} in the WKB theory),: for $\la \sim \mu$ the solutions
have an Airy asymptotics.
However the specific equation (\ref{eq:modstationary2}) is rather non-standard because, as we know form the variational analysis,
the turning point $\mu$ is approaching $1$ as $D \to 0$ and thus it is merging with the singularity $\la = 1$. The merging of the Bessel and Airy
asymptotics yields, as we will shortly show, an asymptotic governed by a Schr\"odinger equation with an Airy potential plus a centrifugal term.

{\em Asymptotic for $\la \sim 0$} \\
We start by computing the small $D$ limit of (\ref{eq:modstationary2}) for $\la \sim 0$.
To this aim we make the transformation
$w=D^{-\frac12}\la$, $y_0(w)=p(D^{\frac12}w)$ to get
\begin{equation*}
 y_0''(w)-\frac bw y_0'(w)+(-1+\frac b{w^2})y(w)=0+ O(D^{\frac12}) \; .
\end{equation*}
Disregarding the higher order term the general solution of the latter equation \cite{bateman2} is
$$y_0(w)=z^{\frac{1+b}{2}}\big( \alpha_0 I(\frac{b-1}{2},w) +  \beta_0 I(-\frac{b-1}{2},w) \big)$$
where $I(\nu,w)$ is the modified Bessel function of the first kind of order $\nu$. Since $I(\nu,w) \propto w^{\nu}$
as $w \to 0$ and  $y_0(w) \propto w^b$, we deduce that the correct behaviour
is given by $\beta_0=0$. Inserting the original variable (and relabelling the constant $\alpha_0$), we deduce that
\begin{equation}\label{eq:at0App}
p(\la)\sim D^{-\frac{b-1}{4}}\alpha_0 \la^{\frac{1+b}{2}}I(\frac{b-1}{2}, D^{-\frac12} \la)
\end{equation} 
for some $\alpha_0>0$ to be computed.

{\em WKB bulk asymptotic}\\

In the bulk, that is for $0<\la<\mu$, the solutions of (\ref{eq:modstationary}) obey the standard WKB asymptotic, namely
$p(\la) \sim y_B(\la) = R(\la) e^{\pm D^{-\frac12} S(\la) } $ for some rational function $R(\la)$. An easy computation yields
\begin{equation*}
 y_B(\la) \sim  \la^{\frac b2} (1-\la)^{\frac a2-\frac14} \, \big( \alpha_B
 e^{- D^{-\frac12} \frac{2(1-\la)^{\frac32}}{3}}+
 \beta_B \, e^{+ D^{-\frac12} \frac{2(1-\la)^{\frac32}}{3}} \big) \; .
\end{equation*}

In order to compute the parameters $\alpha_B,\beta_B$ we must \textit{match} the WKB and Bessel asymptotic
in that transition region where both are valid, namely for 
$\la = O(D^{\gamma})$ for $0<\gamma <\frac12$. This boils down to compare the $\la \to 0$ expansion of
the WKB asymptotic with the $w \to \infty$ expansion of the
Bessel asymptotic

Computing the small $\la$ expansion of the WKB asymptotic one gets
\begin{equation*}
 y_B(\la) \sim  \la^{\frac b2}  \, \big( \alpha_B e^{- \frac23 D^{\frac12}}
 e^{+ D^{-\frac12} \la }+
 \beta_B \, e^{+ \frac23 D^{\frac12}}
 e^{- D^{-\frac12} \la } \big) \; .
\end{equation*}
On the other side, the modified Bessel function has the well known \cite{bateman2} behaviour at infinity
$I(\nu,w)\sim \frac{e^w}{\sqrt{2\pi w}}$: Using the latter formula and (\ref{eq:at0App}) one gets
\begin{equation*}
 p(\la)\sim \alpha_0\la^{\frac{b}{2}}e^{D^{-\frac12}\la}
\end{equation*}
We conclude that the matching yields $\beta_B=0$ and $\alpha_0=D^{\frac{b-1}{4}} e^{- \frac23 D^{\frac12}}\alpha_B$. Then in the bulk
the steady-state solution has the asymptotic
\begin{equation}\label{eq:atbulkApp}
  p_s(\la) \sim \alpha_B   \la^{\frac b2} (1-\la)^{\frac a2-\frac14}
 e^{- D^{-\frac12} \frac{2(1-\la)^{\frac32}}{3}}
\end{equation}
for some $\alpha_B >0$ to be computed.

{\em The asymptotic for $\la \sim \mu \sim 1$}\\

As we know from the variation estimate $\mu=1-O(D^{\frac13})$. Therefore, without losing in generality, we can assume that
$\mu=1-A D^{\frac13} + o(D^{\frac13})$, with $A \geq0$.

In order to analyse the steady state in the region $\la \sim 1$, we re scale the variable $w(\la)=D^{-\frac{1}{3}}(1-\la)$ and the steady-state
\begin{equation}\label{eq:at1App}
 p_s(\la)=D^{-\frac{1}{3}} w^{\frac{a}{2}}y_1(w) \big( 1 + O(D^{\frac12})  \big)\; .
\end{equation}
With this definition the rescaled distribution $y_1(w)$ is easily seen to satisfy
the Airy-like differential equation
\begin{equation}\label{eq:diffmodelApp}
 y''(w)=\big( w-A + \frac{a(a-2)}{4 w^2} \big) y(w) \;, \qquad w\geq 0 \; .
\end{equation}
Close to $w \to 0$ the solutions of the above equation behaves as
$y_1 \propto w^{\rho} $ with $\rho=\pm \frac a2$. Since $p_s \sim (1-\la)^a$, we must impose
the non trivial condition $y_1 \propto w^\frac a2$.

In order to further characterise the solution $y_1$ of (\ref{eq:diffmodelApp}) we
we must match the asymptotic at $\la \sim 1 $ with the bulk asymptotic (\ref{eq:atbulkApp}) in a transition region where both approximation are
feasible, namely for  $1-\la=O(D^\gamma)$ with $0<\gamma<\frac13$. Similarly to the previous matching of asymptotics,
this is done by comparing the $w \to \infty$ limit of (\ref{eq:at1App}) to the $\la \to 1$ limit of the bulk behaviour (\ref{eq:atbulkApp}).
From the standard (WKB) asymptotic analysis - see \cite{fedoryuk93} - of equation (\ref{eq:diffmodelApp}) we know that our solution has the following asymptotics
for $w\gg0$
\begin{equation}\label{eq:alphaApp}
 y_1(w) \sim  w^{-\frac 14}\big( \alpha_1 e^{-\frac{2 w^{\frac 32}}{3}}+ \beta_1 e^{-\frac{2 w^{\frac 32}}{3}}\big)  \;.
\end{equation}
where $\alpha_1,\beta_1$ are constants depending on the parameters $a,A$ and on the normalisation at $w=0$.

Inserting the original variables $p$ and $\la$, the above asymptotics reads
\begin{equation*}
 p(\la) \sim D^{-\frac{a-1/2}6}(1-\la)^{\frac a2-\frac14}  \big( \alpha_1 e^{- D^{-\frac12}\frac{2 (1-\la)^{\frac32}}{3}} +
 \beta_1  e^{+ D^{-\frac12}\frac{2 (1-\la)^{\frac32}}{3}}  \big)
\end{equation*}
Similarly the limit $\la \to 1$ of the bulk asymptotics (\ref{eq:atbulkApp}) reads
\begin{equation*}
  p(\la) \sim \alpha_B  (1-\la)^{\frac a2-\frac14}
 e^{- D^{-\frac12} \frac{2(1-\la)^{\frac32}}{3}}  \;.
\end{equation*}
By the matching procedures we conclude that $\alpha_B=   \alpha_1 D^{\frac{a-1/2}6} $ and $\beta_1=0$.
Namely $y_1(w)$ decays as $w \to \infty$.

We have thus proven that the function $y_1(w)$ satisfies the properties (i,ii,iii) of the
function $Y_a(w)$ that we had introduced in the main body of the paper - see the discussion below equation (\ref{eq:at1}).
Therefore $y_1(w)$ coincides
-up to the normalisation- to $Y_a(w)$. 
Moreover, since our computation of the factors $\alpha_0,\alpha_B$ shows that $p_s$ is asymptotically suppressed for $\la <1$, then
$$\int_0^1p(\la) \sim \int_0^{\infty} w^{\frac a2}y_1(w) dw \; . $$
We deduce that $\int_0^{\infty}  w^{\frac a2}y_1(w) dw=1$ and thus $y_1$ actually coincides with $Y_a(w)$.

Summing up the result of our asymptotic computations, we can give a uniform asymptotic description of the steady state $p_s$ in the whole
unit interval:
\begin{align}\label{eq:at1def}
& p_s(\la) \sim D^{-\frac{1+\frac a2}{3}} (1-\la)^{\frac{a}{2}}Y_a\big(D^{-\frac13}(1-\la) \big) \; , \quad 1-\la=O (D^{\gamma}), \gamma >0 \\  
\nonumber
&  p(\la) \sim \alpha D^{\frac{a-\frac12}6} \la^{\frac b2} (1-\la)^{\frac a2-\frac14}
 e^{- D^{-\frac12} \frac{2(1-\la)^{\frac32}}{3}} \, , \mbox{ for } \lambda \mbox{ in the bulk:} \\ \label{eq:atbulk}
&  O(D^{\gamma_1}) <\la < 1 - O(D^{\gamma_2}), \mbox{ with }\gamma_1<\frac12 , \gamma_2<\frac 13 \\ \label{eq:at0}
& p(\la)\sim \alpha D^{\frac{a-\frac12}6} e^{-\frac{2 D^{-\frac12}}{3 }}\la^{\frac{1+b}{2}}I(\frac{b-1}{2}, D^{-\frac12} \la)
 \; , \quad \la = O(D^{\gamma}), \gamma>0 
\end{align}
Here $I(\cdot,\cdot)$ is the modified Bessel function of the first kind and
$\alpha>0$ is defined by the asymptotic expansion of $Y_a(w)$ for large $w$ 
\begin{equation*}
 Y_a(w) \sim \alpha w^{-\frac 14}e^{-\frac{2 w^{\frac 32}}{3}} \mbox{ for } w \mbox{ large} \;.
\end{equation*}

{

{\em Momenta of the distribution}\\
Not only the mean $\mla$, but all higher momenta of the steady state can be computed from $Y_a(w)$.
In fact, as our computations show, see in particular (\ref{eq:atbulk},\ref{eq:atbulk}),
the steady state $p_s$ is exponentially suppressed away from $\la \sim 1$.
Therefore, using the change of variable (\ref{eq:at1}), we see that the following asymptotic identity holds
\begin{equation}\label{eq:highermomenta}
 \int_0^1(\la-\mla)^k p_s(\la) \sim D^{\frac k3}\int_0^{\infty} (w-\langle w \rangle)^k w^{\frac a2} Y_a(w) dw  
\end{equation}
where $ \langle w \rangle = A$, as can be seen integrating by part equation (\ref{eq:diffmodel}) and using the
normalisation $\int_0^1w^{\frac a2} Y_a(w) dw =1$.

\subsubsection*{Relaxation time}

Here we briefly analyse the asymptotics of the linear stability equation   
\begin{equation*}
 - \eta \delta(\la)= (\la -\mla_s) \delta - p_s(\la) \int_0^1 \la \delta(\la) d \la +
D \left[\frac{\partial^2 p}{\partial\lambda^2} -\frac{\partial}{\partial\lambda} \left[p(\lambda)
\frac{\partial}{\partial\lambda}(\log q(\lambda))\right]\right] \; ,
\end{equation*}
where $\mla_s=\int_0^1 \la p_s(\la) d\la$.

Since, as we have already discussed, in the small $D$ regime $p_s(\la)$ is localised at $\la \sim 1$ (and exponentially suppressed for $\la < 1$),
then $\delta$ too must be localised at $\la \sim 1$. Therefore we can analyse $\delta$
using the same scaling analysis that we used to analyse $p_s(\la)$ - see equation (\ref{eq:at1App}):
Defining $w=D^{\frac{1}{3}}(1-\la)$, $1-\mla_s=AD^{\frac13}+o (D^{\frac 13})$ and $T=w^{\frac a2}\delta(w) \big( 1 + o(1) \big) $,
we find that the linearised equation for
the perturbation $\delta$ has a well-defined limit if and only if $\eta=\eta_a D^{\frac13} +o(D^{\frac13})$, and the limit is
\begin{equation}\label{eq:tauat1}
 \eta_a T(w)=- T''(w) +\big( w-A + \frac{a (a-2)}{4 w^2} \big) T(w)+ Y_a(w) \int_0^{\infty}w^{1+\frac a2} T(w) dw  \; .
\end{equation}
Here $\eta_a=\lambda_1-A$ where $A$ is the ground-state energy of the Airy like Hamiltonian $H$ (\ref{eq:AiryHamiltonian}) and $\lambda_1$
the energy of the first excitation.

In order to show that $\eta_a=\lambda_1-A$, we notice that 
that $T(w)$ must be an eigenvector of the following operator
\begin{equation*}
 H_T= -\frac{d^2}{dw^2}+ (w-A) +\frac{a(a-2)}{4 w^2} + Y_a(w)\int_0^{\infty} dw w^{1+\frac a2} \cdot     	\;,
\end{equation*}
where $\int_0^{\infty} dw w^{1+\frac a2} \cdot$ applied to a function $f$ is simply $\int_0^{\infty} f(w) w^{1+\frac a2} d w$.
Moreover $T(w)$ must satisfy the same boundary conditions as $Y_a(w)$, namely
\begin{itemize}
 \item[(i)]$T(w) \propto w^{\frac{a}{2}}$ as $w \to 0$.
 \item[(ii)]$\lim_{w \to +\infty} T(w)=0$
\end{itemize}
which reflects the analogous properties of $\delta$, respectively
$\delta \propto (1-\la)^a$ and $\delta$ is localised at $\la \sim 1$. 

The spectrum of the operator $H_T$ is of the form $\lambda^{(k)}-A$, where
$\lambda^{(k)}$ is the $k-th$ excited state of the Airy-like Hamiltonian H (\ref{eq:AiryHamiltonian}). Therefore the lowest
eigenvalue is $\lambda_1-A$. Indeed, we define
\begin{equation*}
 \delta^{(k)} (w) = Y_a(w)+Y^{(k)}(w)
\end{equation*}
where $Y^{(k)}$ is the $k-th$ excited eigenstate of $H$ normalised by the formula $\int_0^{\infty}w^{1+\frac a2}Y^{(k)}(w) dw=\lambda^{(k)}- 2 A$.
Then an easy computation - recall $\int_0^{\infty}w^{1+\frac a2}Y_a(w) dw=A$ - shows that $H_T[\delta^{(k)}]=(\lambda^{(k)}-A)\delta^{(k)}$.

We finally remark that the condition
$\int \delta d \la=0$ is (asymptotically) equivalent to the condition $\int_0^{\infty} w^{\frac a2}T(w) dw=0$
which is a direct consequence of (\ref{eq:tauat1}) and $\eta_a \neq 0$.

}
\subsection*{Numerics}

Upon passing to $\phi$ such that $p=q \exp (\phi)$, the equation for $\phi$ is
\begin{equation}
\dot{\phi} = \lambda-\langle \lambda \rangle + D\left(  ( (\log q)'+\phi' )\phi'+\phi'' \right) 
\end{equation}
where the prime and the dot stand for derivation with respect to $\lambda$ and $t$, respectively.  We divide the interval $[0,1]$ in $N$ equal parts of length $h=1/N$ and consider the discrete equation, upon considering discrete evaluation of the  derivatives:
\begin{eqnarray}
\phi_i' &=& \frac{-\phi_{i+2}+8\phi_{i+1}-8\phi_{i-1}+\phi_{i-2}}{12h} \quad 2 \leq i \leq N-3 \nonumber \\
\phi_1' &=&\frac{\phi_2-\phi_0}{2h} \quad \phi_0' = \frac{\phi_1-\phi_0}{h} \nonumber \\
\phi_{N-2}' &=&\frac{\phi_{N-1}-\phi_{N-3}}{2h} \quad \phi_{N-1}' =\frac{\phi_{N-1}-\phi_{N-2}}{h} 
\end{eqnarray}
and analogous formula for $\phi''$. The discrete time evolution equations read
\begin{eqnarray}
\phi_{i,t+dt} =\phi_{i,t}+f_i dt \nonumber \\
f_i = i/N-\langle \lambda \rangle +D\left(  ( (\log q)_i'+\phi_i' )\phi_i'+\phi_i'' \right) 
\end{eqnarray}
This forward simple scheme is numerically stable and convergent if $dt<\frac{1}{2} h^2$\cite{crank1979mathematics}.
{In any case we performed simulations at fixed $D$ and calculated $\langle \lambda \rangle$ and $\langle \sigma \rangle$ from the distribution (stationary and/or variable in time).}
The relaxation time has been calculated numerically by keeping track during a simulation of the $L^{1}$ distance in time between the current distribution and the stationary one,
\begin{equation}
d(t) = \int d\lambda |p(\lambda,t)-p_s(\lambda)|,
\end{equation}
starting from a perturbed state of the form $p(\lambda,t=0) \propto p_s(\lambda) e^{\epsilon \lambda}$, where $|\epsilon|<<1$. The decay of the distance in time is well fitted by an exponential law 
\begin{equation}
d(t) \simeq e^{-t/\tau}
\end{equation} 
with a decay constant $1/\tau$ that depends on $D$ alone, as soon as  $|\epsilon|<<1$.

\section*{Acknowledgements}
 D. De Martino is supported by the People Programme (Marie Curie Actions) of the European Union's Seventh Framework Programme (FP7/2007-2013) under REA grant agreement no. $[291734]$. D. Masoero is supported by the
FCT scholarship, number SFRH/BPD/75908/2011.  D. De Martino thanks the Grupo de F\'isica Matem\'atica of the Universidade de Lisboa
for the kind hospitality. We also wish to thank Matteo Osella, Vincenzo Vitagliano and Vera Luz Masoero for useful discussions, alaso late at night.

\bibliographystyle{unsrt}

\section*{References}
\bibliography{reference}

\end{document}